\documentclass[runningheads]{llncs}
\usepackage{cite}
\usepackage{graphicx}
\usepackage{array}
\usepackage{makeidx}  
\usepackage{multirow}
\usepackage{slashbox}
\usepackage{algorithm}
\usepackage{algorithmic}
\usepackage{amsmath}
\usepackage{xcolor}
\usepackage{subfig}
\usepackage{comment}

\begin{document}
\title{Data Management in Time-Domain Astronomy: Requirements and Challenges}
\titlerunning{Data Management in Time-Domain Astronomy}
%
\author{Chen Yang\inst{1} \and Xiaofeng Meng\inst{1}\thanks{Corresponding author is Xiaofeng Meng} \and Zhihui Du\inst{2} \and Zhiqiang Duan\inst{1} \and Yongjie Du\inst{1}}
\authorrunning{Chen Yang et al.}
%
\institute{School of Information, Renmin University, China \and
Department of Computer Science and Technology, Tsinghua University, China}
\maketitle              
\hyphenation{space straightforward storage}
\begin{abstract}
In time-domain astronomy, we need to use the relational database to manage star catalog data. With the development of sky survey technology, the size of star catalog data is larger, and the speed of data generation is faster. So, in this paper, we make a systematic and comprehensive introduction to process the data in time-domain astronomy, and valuable research questions are detailed. Then, we list candidate systems usually used in astronomy and point out the advantages and disadvantages of these systems. In addition, we present the key techniques needed to deal with astronomical data. Finally, we summarize the challenges faced by the design of our database prototype.
\end{abstract}
\section{Introduction}
The origin of information explosion is astronomy, which is first facing challenges of big data\cite{Naimi2014Big}. In the new century, with the development of astronomical observation technique, astronomy has already entered a informative big data era and astronomical data is rapid growth in terabytes (TB) or even petabytes (PB). When Sloan Digital Sky Survey project started in 2000, data being collected by telescope in New Mexico in few weeks is greater than all historical data. In 2010, information files contained 1.4 $\times$ 2$^{42}$ bytes. But, Large Synoptic Survey Telescope (LSST) can be used in Chile in 2019, which can get the same information within 5 days. Now, a number of countries are running large-scale sky survey project. Except SDSS, these projects include PanSTARRS (Panoramic Survey Telescope and Rapid Response System), WISE (Widefield Infrared Survey Explorer), 2MASS (Two Micron All Sky Survey), Gaia of the European Space Agency (ESA), UKIDSS (UKIRT Infrared Deep Sky Survey), NVSS (NRAO VLA Sky Survey), FIRST (Faint Images of the Radio Sky at Twenty-cm), 2df (Two-degree-Field Galaxy Redshift Survey), LAMOST (Large Sky Area Multi-Object Fiber Spectroscopic Telescope), GWAC (Ground Wide Angle Camera) in China and so on. These sky surveys are generating a large number of astronomical data.

Astronomical data also has four V characteristics of big data: Volume, Velocity, Variety and Veracity\cite{Bryant2008Bigdata}. For example, LSST covers all sky and store in the database one cycle per 7 days. GWAC covers 5000 degree$^2$ and stores in real-time per 15 seconds. Astronomy moves on to a big data era. Therefore, it is important to study big data which is generated by astronomy\cite{Cui2015Big}. However, the existing database systems cannot support the demand of astronomical data, especially for the real-time and GWAC's scalability. We need to design our database system for the discovery of transient celestial phenomena in short-timescale.

The rest of the paper is organized as follows. Section \ref{section:backbround} surveys the background. Then, there are the problem definition and basic knowledge in Section \ref{section:preliminaries}. And, in Section \ref{section:candidateSystems}, we list candidate systems used in astronomy and point out the advantages and disadvantages of these systems. Then, in Section \ref{section:functionalAnalysis}, we give the functional analysis of our database prototype. In addition, we point out the challenges to the design of our database prototype in Section \ref{section:mainChallenge}. Finally, Section \ref{section:Summary} concludes this paper.

\section{Background}\label{section:backbround}
GWAC is built in China, which consists of 36 wide angle telescopes with 18cm aperture. Each telescope equips 4k $\times$ 4k charge coupled device (CCD) detector. Cameras cover 5000 degree$^2$. Temporal sampling is 15 seconds. Cameras detect objects of fixed sky area for 8 hours each observation night. GWAC has special adventure in the time-domain astronomical observation, according to size of observation field and sampling frequency of observation time. It is great challenges for data management and processing in giant data and high temporal sampling.

\begin{table*}[!hbt]
\caption{Dada volume generated by GWAC}\label{tab:DadavolumegeneratedbyGWAC}
\centering
\begin{tabular}{|c|c|c|c|c|c|c|}
\hline
\multirow{2}*{Cameras} & \multicolumn{2}{|c|}{One Day(8 hours)} & \multicolumn{2}{|c|}{One Year(260 days)} & \multicolumn{2}{|c|}{Ten Years}\\
\cline{2-7}
&Records & Size & Records & Size &Records & Size\\
\hline
1 & 3.37 $\times$ 10$^8$ & 61.88GB & 8.77 $\times$ 10$^{10}$ & 15.71TB  & 8.77 $\times$ 10$^{11}$ & 157.1TB \\
\hline
36 & 1.21 $\times$ 10$^{10}$  & 2.17TB & 3.16 $\times$ 10$^{12}$ & 565.62TB & 3.16 $\times$ 10$^{13}$ & 5.52PB \\
\hline
\end{tabular}
\end{table*}
As shown in Table \ref{tab:DadavolumegeneratedbyGWAC}, GWAC works 8 hours a night, 260 days a year on average. The index of star catalog data in GWAC includes: each star catalog in one image having 1.756 $\times$ 10$^5$ records. So, camera array can generate 6.3 $\times$ 10$^6$ records in 15 seconds, contain 1920 $\times$ 36 = 69120 images and occupy about 2.17TB storage space in each night. The requirements of database management systems (DBMS) are:  (1) rapid big data storage capacity that all star catalogs are ingested within 15 seconds and 2.17TB star catalog data in each observation night should be stored before next observation night. (2) high-speed data acquisition can be analyzed in real-time and rapid contextual computing capacity when facing mass of incessancy and high-density star catalogs. This means relevance star catalog data generated by one CCD within 15 seconds and reference star catalog to form light curves. (3) In 10-year design cycle, GWAC will generate about 5.53PB size of star catalogs. Therefore, DBMS for storing star catalog data must have the great management ability for massive data.

For GWAC, the most immediate way to data management and design of processing system is the database (only for data storage) and peripheral program (rapid operation and the result obtaining). Xu et al. \cite{Xu2013A} researched by cross-match for key technique developed a sky partitioning algorithm based on longitude and latitude of space and increased the speed of cross match compute rapidly. Based on the advantage of parallel compute by graphics processor, Zhao et al. \cite{Zhao2013Accelerating} used graphics processor accelerate method to speed up the image subtraction processing. Zhao et al. \cite{Zhao2013Parallelizing} developed point source extracting program SEXtractor in the field of astronomy which is developed by graphic processor accelerate. Wang et al. \cite{Wang2013Accelerating} developed a cross-match accelerating algorithm based on graphic processor. The advantages of this plan are the straightforward thoughts and many mature techniques. The disadvantages are that database is exchanging with peripheral programs and bring useless time loss of I/O. Combination of program results in lack of optimizations as a whole.

Jim Gray directed the development of Skyserver and purposed Zone algorithm \cite{zone1,zone2}. It means that we can use SQL of DBMS to realize hyperspace index replacement classical Hierarchical Triangular Mesh (HTM). This method can reduce data interaction and increase speed. This is the principle of large-scale scientific computation and database architecture design: designing philosophy to bring computation to data rather than putting data to computation\cite{Szalay2009Gray}. This paper inspired by this idea, purposes design idea that combines data processing of GWAC and data management to a database platform.

Massive astronomical data is great challenge for data storage and management. Therefore, rapid processing of massive astronomical data is very important. GWAC astronomical database can provide inquiry service and form a light curve. The database contains two scientific targets:

\begin{itemize}
  \item \textbf{Rapid big data storage capacity}. All star catalogs generated by cameras can be stored within 15 seconds, and 2.17TB star catalog data in each observation night should be stored before next observation night.
  \item \textbf{High-speed data collection}. Data can be analyzed in real-time, including efficient detection and dynamic recognition astronomical object.
\end{itemize}

The main scientific target of GWAC is to search optical transient sources in real time and locate in observation sky and formulate star catalog index. GWAC works sky survey every 15 seconds. Facing incessancy observation intensive stellar field and massive star catalog in short timescale, data processing system must have relevance computing ability to rapidly recognize celestial objects and data processing algorithm, meaning relevance star catalog data generated by each CCD in 15 seconds and reference star catalog to generate light curve. So, the goal is that we need to develop a database which can integrate the algorithm of the point source identification and has high expansibility.

\section{Preliminaries}\label{section:preliminaries}
\subsection{Problem Definition}
\textbf{Point source extraction}. The camera array ($\Omega$) is consist of 36 wide angle telescopes. Each wide angle telescope T$_i$(i=1, 2, ..., 36) equips 4k $\times$ 4k CCD detector. Point source extraction is to transform optical image into figure signal by CCD detector which forms star catalog data.

Each row of data in star catalog is used to record one star. Property information\cite{Meng2016Column} for each star shows on Table \ref{tab:starcatalog}. Each image in star catalog has about 1.756 $\times $10$^5$ records.

\begin{table*}[!hbt]
\caption{Star catalog attributes of each source}\label{tab:starcatalog}
\centering
\begin{tabular}{|c|c|p{9cm}|}
\hline
Name & Type & Description\\
\hline
ID & long int & Every inserted source measurement gets a unique id. generated by the source extraction procedure.\\
\hline
imageid & int & The reference ID to the image from which the source was extracted.\\
\hline
zone & small int & The zone ID in which a source declination resides, calculated by the source extraction procedure.\\
\hline
ra & double & Right ascension of a source (J2000 degrees), calculated by the source extraction procedure.\\
\hline
dec & double & Declination of a source (J2000 degrees) as above.\\
\hline
mag & double & The magnitude of a source.\\
\hline
mag\_error & double & The error of magnitude.\\
\hline
pixel\_x & double & The instrumental position of s source on CCD along x.\\
\hline
pixel\_y & double & The instrumental position of s source on CCD along y.\\
\hline
ra\_err & double & The 1-sigma error on ra (degrees).\\
\hline
dec\_err & double & The 1-sigma error on declination (degrees).\\
\hline
x & double & Cartesian coordinates representation of RA and declination, calculated
by the source extractor procedure. \\
\hline
y & double & Cartesian coordinates representation of RA and declination, as above.\\
\hline
z & double & Cartesian coordinates representation of RA and declination, as above.\\
\hline
flux & double & The flux measurements of a source, calculated from the mag value.\\
\hline
flux\_err & double & The flux error of a source.\\
\hline
calmag & double & Calibrated mag.\\
\hline
flag & int & The source extraction uses a flag for a source to tell for instance if an
object has been truncated at the edge of the image.\\
\hline
background & double & The source extraction estimates the background of the image.\\
\hline
threshold & double & The threshold indicates the level from which the source extraction should
start treating pixels as if they were part of objects.\\
\hline
ellipticity & double & Ellipticity is how stretched the object is.\\
\hline
class\_star & double & The source extractions classification of the objects.\\
\hline
\end{tabular}
\end{table*}
\begin{figure}[H]
\centering
\includegraphics[width=.6\textwidth]{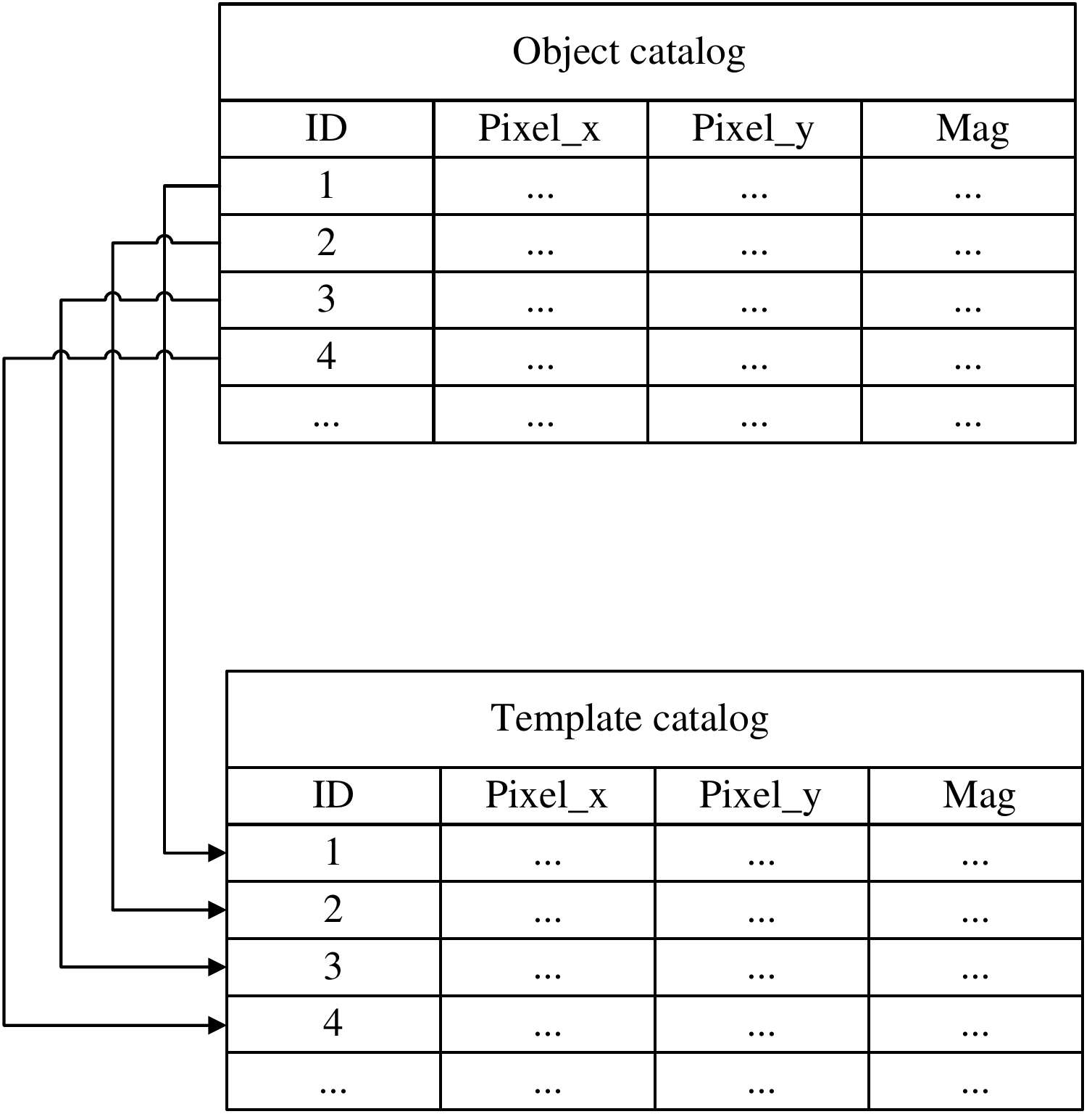}
\caption{Example of cross-match} \label{fig:crossmatch}
\end{figure}
\textbf{Cross-match}. It contrasts and matches the object catalog with template catalog. As shown in Figure \ref{fig:crossmatch}, if object catalog could match template catalog, the pipeline will enter timing sequence photometry channel to process and manage light curve. If the star cannot be matched in template catalog, it is transient source (candidate). Cross-match is the key algorithm in GWAC searching transient source and generation light curve. Cross-match issue must depend on effective partition strategy. We will divide the sky into each horizontal strip in the pixel coordinate and each source has a strip belonging to itself. At first, cross-match can compare strip property and decrease times of comparing. Strip can be integrated inside database as the basic unit of data processing.

\textbf{Light curve}. It is the change of the brightness of the object relative to the time. It is the function of time, which usually shows a particular frequency interval.

\subsection{Camera Array Processing Flow}
\begin{figure}[H]
\centering
\includegraphics[width=.6\textwidth]{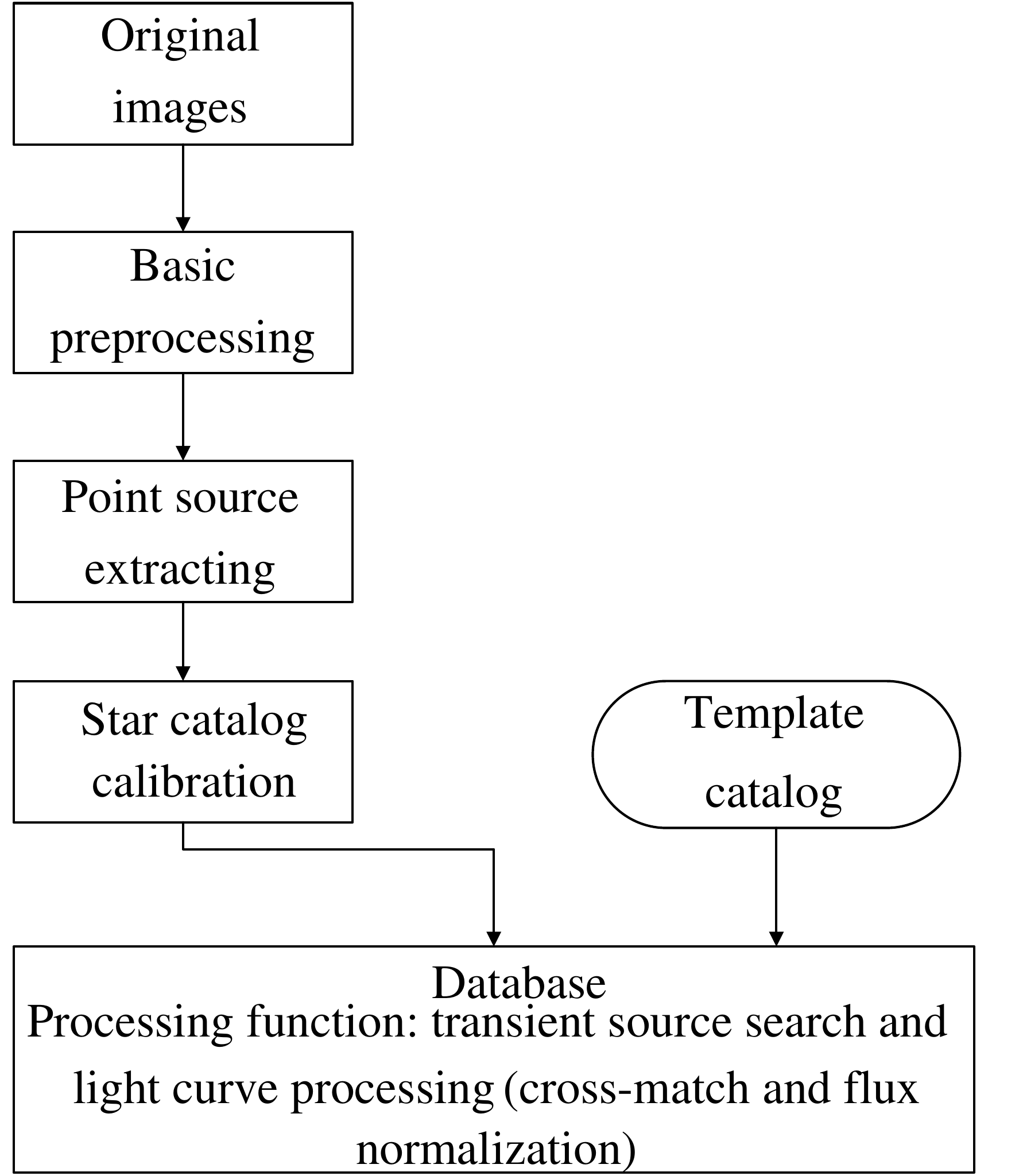}
\caption{GWAC data processing}\label{fig:dataprocessing}
\end{figure}

There is a flow graph for GWAC data processing in Figure \ref{fig:dataprocessing}. According to basic preprocessing of original image, it will extract point source and astrometric calibration of star catalog. Then, it completes tasks about relative discharge calibration, real-time dynamic identification of astronomical object and light curve mining based on cross-match observation data and reference star catalog.

Overall, objectives of our database prototype are: (1) capacity of rapid storage data, object catalogs generated by cameras should be ingested within 15 seconds. 2.17TB star catalog data in each observation night should be stored before next observation night. (2) high-speed data acquisition can be analyzed in real time and has rapid contextual computing capacity when facing mass of incessancy and high-density star catalog. This means relevance star catalog data generated by one CCD in 15 seconds and reference star catalog to generate light curve.

\section{Functional Analysis \& Requirements}\label{section:functionalAnalysis}

\begin{figure}[H]
\centering
\includegraphics[width=.7\textwidth]{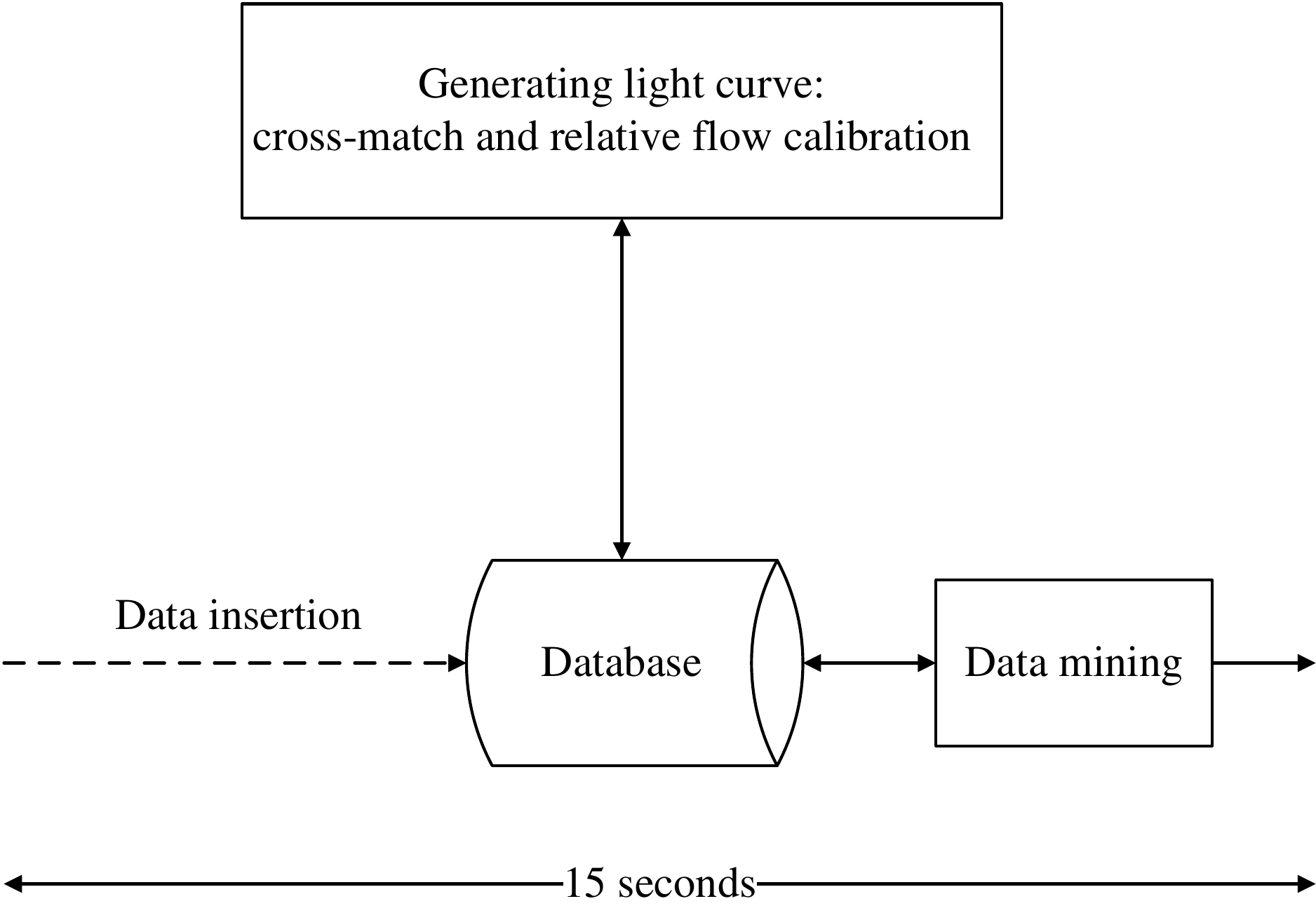}
\caption{The goal of core processing}\label{fig:coreprocessing}
\end{figure}

The key steps in processing GWAC's data are shown in Figure \ref{fig:coreprocessing}. Within 15 seconds, we need to ingest 1.756 $\times$ 10$^5$ records into the database, complete the cross-match, generate the light curve and fulfill the task of data mining. The core functional analysis is as follows:

\subsection{Real-Time Storage}
The system could rapidly store object star catalog generated by cameras in 15 seconds. 2.17TB star catalog can be stored as increment data and make sure that data could be stored in real time. Also, system merges increment data daily during data storage low cycle (non observation night and other time) to increase data storage capacity and decrease storage delay.

\subsection{Cross-match}
The system should establish efficient index mechanism, optimize database join operation, increase star catalog relevance and efficiency of cross-match.

\subsection{High Scalability}
As the observation data of each wide angle telescope is increasing, it is hard to use one server storage and analysis different telescopes data. We need to design high reliable distributed clusters architecture. There are different subset clusters for every wide angle telescope storage star catalog data to ensure consistency of star catalog data. Also it realize whole system processing capacity linear growth and high throughput rate and low delay.

\subsection{Data Mining}
\begin{figure}[H]
\centering
\includegraphics[width=.7\textwidth]{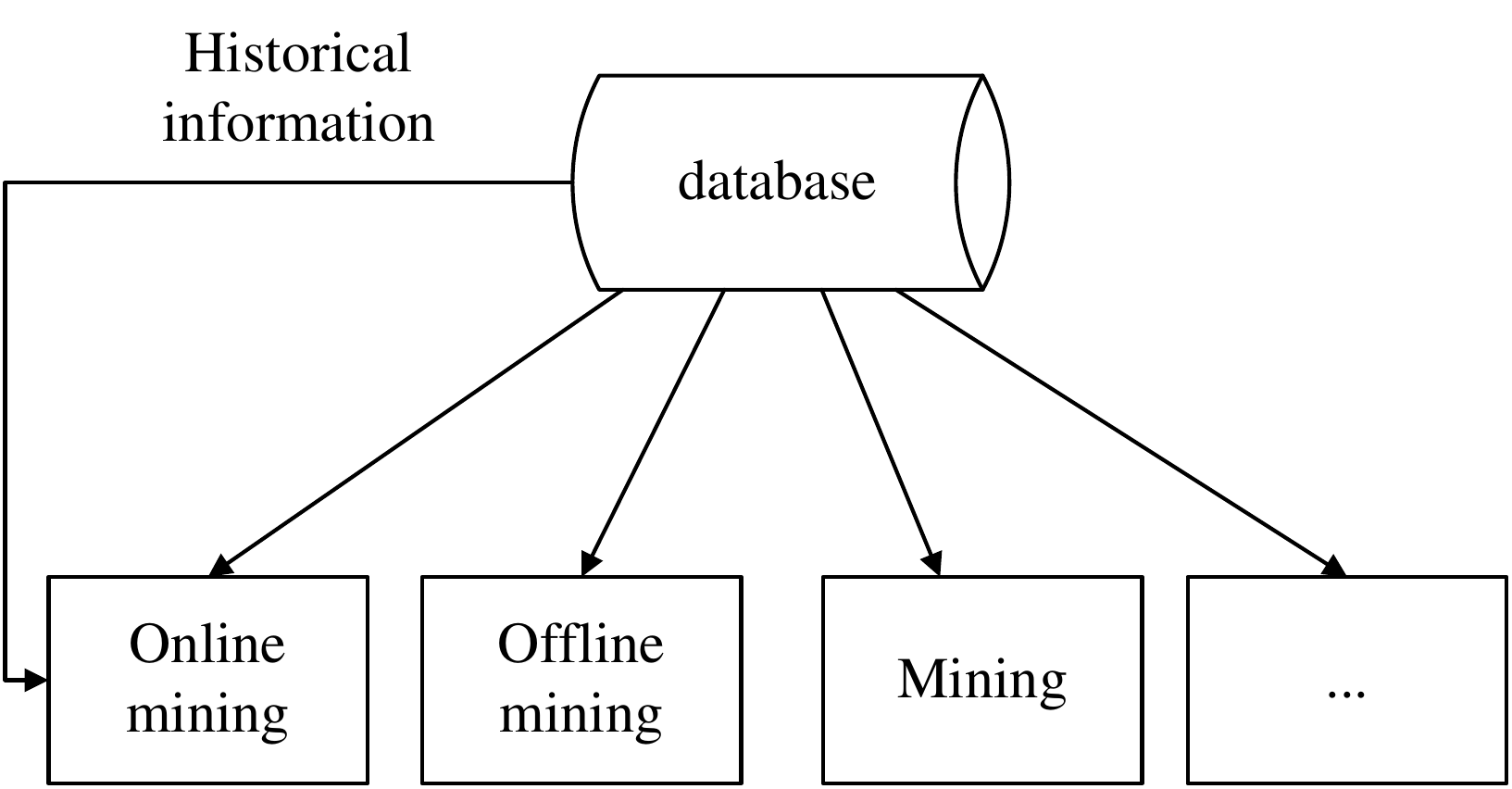}
\caption{Data mining}\label{fig:datamining}
\end{figure}

In database, we need to use technique of data mining to find meaningful astronomical phenomena. The process of data mining can divide into online mining and offline mining (shown in Figure \ref{fig:datamining}).

\textbf{Online mining}. For data flow which observation time is shorter than one night, real-time monitoring and inner analysis windows should be used for dynamic recognition of astronomical targets.

\textbf{Offline mining}. For long-timescale data which observation time is longer than one night, using full scale historical data predict waveform of light curve and judging change cycle of curve to analyze fluctuation features of star body is better.

\section{Major Challenges}\label{section:mainChallenge}
Based on the functional analysis in Section \ref{section:functionalAnalysis}, the main challenge issues for our database prototype can be concluded as below.

\subsection{Customized Operators}
For characteristics of astronomical data, we can customize operators in our database prototype. There are the customized operators as below:
\begin{itemize}
  \item Increment storage operator ``DeltaInsert" ensures data real-time storage.
  \item Range join operator ``RangeJoin" ensures rapid cross-match.
  \item Data mining operator library ensures association analysis, classify, cluster, outliers detector and other functions.
\end{itemize}

In addition, our database prototype is designed with batch processing and query plan adapter of streaming processing, by uniform query system interface. it cannot only get real-time data flow query result, but also it can get query result of full-history form of offline historical data.
\begin{table*}[h]
\caption{Different system characteristics}\label{tab:Differentsystemcharacteristics}
\centering
\begin{tabular}{|p{1.8cm}|p{2cm}|p{2cm}|p{2cm}|p{2cm}|p{2cm}|}
\hline
Features & SciDB & OceanBase & MonetDB & MongoDB & Spark\\
\hline
Storage engine & Array data model & Memory affair storage & Binary association table & Assembly storage & RDD \\
\hline
Advantage & Shared-nothing design, SciDB-R & Support ACID Affair, Fault tolerant automatically and load balancing, Division storage & Automatic indexing, index not taking extra storage space & Powerful query language, support dynamic query and fully index & Easy to scale out clusters and
fast iterative computations\\
\hline
Disadvantage & Not support complex search condition & Low open source version & Low speed of insert operation & Not support transaction operation & Low data storage efficiency\\
\hline
\end{tabular}

\end{table*}

\subsection{Large-scale Data Management}
With the increment of the amount of data generated by GWAC, our database prototype needs to deal with the problem of data management in PB level. A large-scale data management engine needs to be designed to ensure data consistency and integrity, and easy to scale out.

\subsection{Scalable Query Processing}
In a large-scale cluster environment, our database prototype needs to ensure a low query response time. We should use the design philosophy of massively parallel processing (MPP) to implement scalable query processing.

\subsection{Long-term Data Storage}
Since the life cycle of GWAC is 10 years, it is essential to provide the hardware and the storage strategy to save all historical data. Original image data can verify correctness of related analysis and provide original data for image analysis in depth.

\section{Candidate Systems}\label{section:candidateSystems}

For characteristic and functional requirements of astronomical data, there are some candidate systems preparing to be compared and their characteristics are summarized in Table \ref{tab:Differentsystemcharacteristics}.

\subsection{SciDB}
SciDB is a new science database for scientific data. Application areas include astronomy, particle physics, fusion, remote sensing, oceanography, and biology \cite{Stonebraker2009Requirements}. Scientific data often does not fit easily into a relational model of data. Searching in a high dimensional space is natural and fast in a multi-dimensional array data model, but often slow and awkward in a relational data model.

Array DBMS is a natural fit for science data. So, SciDB uses array data model as the storage engine. Another characteristic of SciDB is sparse or dense array. Some arrays have a value for every cell in an array structure. It have ability to process skewed data. Moreover, SciDB uses the shared-nothing distributed storage framework. This is easy to scale out the cluster. In addition, SciDB has built an interface for R language that lets R scripts access data residing in SciDB. However, SciDB can not support complex search condition. And, the effect of data storage in real-time is not good.

\subsection{OceanBase}
OceanBase\cite{oceanbase} is a high-performance distributed database. It can realize cross row and cross table affairs based on hundred billions records and hundreds TB data.

Because OceanBase is a relational DBMS, it uses memory affair storage as the storage engine, and can support ACID affair. And, in the distributed environment, OceanBase provides automatic fault tolerance, load balancing and division storage. However, the open source version of OceanBase is low, and the stability of the system is not strong.

\subsection{MonetDB}
MonetDB is also a relational DBMS for high-performance applications in data mining, scientific databases, XML Query, text and multimedia retrieval, that is developed at the CWI database architectures research group since 1993 \cite{Manegold2009Database}.

MonetDB is designed to exploit the large main memory of modern computer systems effectively and efficiently during query processing, while the database is persistently stored on disk. The core architecture of MonetDB has proved to provide efficient support not only for the relational data model and SQL, but also for the non-relational data model, e.g., XML and XQuery \cite{Boncz2006MonetDB, Idreos2012MonetDB}. In addition, MonetDB supports column-store by using binary association table as storage engine.  Moreover, It can build the index automatically, the index can not take extra storage space. Yet, the efficiency of insertion operation is low, especially for incremental insertion.

\subsection{MongoDB}
MongoDB \cite{mongodb} is an agile database that allows schemas to change quickly as applications evolve, while still providing the functionality that developers expect from traditional databases.

\begin{figure*}[!hbt]
\centering
\includegraphics[width=.9\textwidth]{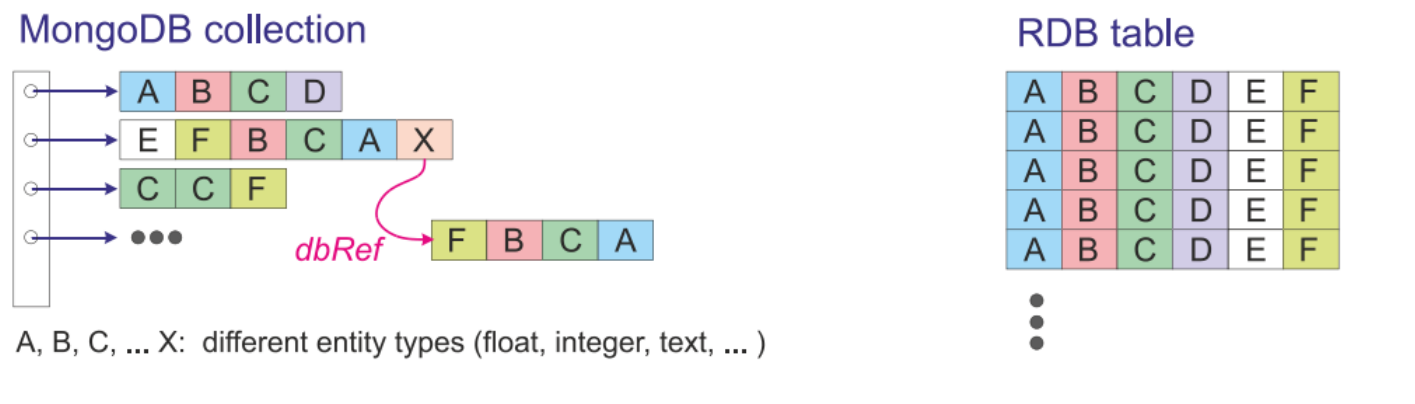}
\caption{Basic data structure between mongoDB and a relational database}\label{fig:mongodb}
\end{figure*}

Figure \ref{fig:mongodb} shows the basic difference between the schema-free document database structure and the relational database\cite{Parinaz2014Management}. While the tables in a relational database have a fixed format and fixed column order, a MongoDB collection can contain entities of different types in any order. The element dbRef allows the creation of an explicit reference to another document in the same database or in another database on another server.

MongoDB uses assembly storage as storage engine to support dynamic query and fully index, and has a powerful query language. However, it can not support transaction operation well.

\subsection{Spark}
Apache Spark \cite{spark} is an open-source cluster computing framework for big data processing. It has the distributed data frame, and goes far beyond batch applications to support a variety of compute-intensive tasks including interactive queries, streaming, machine learning and graph processing \cite{Shanahan2015Large}.

Spark use RDD \cite{Zaharia2012Resilient} as storage engine to ensure the correctness and fault tolerance of the query processing. It is easy to scale out clusters, and has fast iterative computing power. However, because spark needs to rely on other storage frameworks, its data storage efficiency is low.

\section{Summary}\label{section:Summary}
We investigate and survey requirements of databases in astronomy and introduce the background knowledge, points out core problems and the main challenges. However, none of these candidates is suitable for large time-domain surveys and that a new system should be developed to meet the challenges.
\section{Acknowledgement}
This research was partially supported by the grants from the National Key Research and Development Program of China (No. 2016YFB1000602, 2016YFB1000603); the Natural Science Foundation of China (No. 91646203, 61532016, 61532010, 61379050, 61762082);  the Fundamental Research Funds for the Central Universities, the Research Funds of Renmin University (No. 11XNL010); and the Science and Technology Opening up Cooperation project of Henan Province (172106000077).

\bibliographystyle{plain}
\bibliography{sigproc}
\end{document}